\documentclass[12pt]{article}
\usepackage{amsfonts}
\usepackage{amssymb}
\usepackage{color}
\usepackage{graphicx}

\newcommand{\Sc}{{\cal S}}

\newcommand{\mc}{\mathcal}

\newcommand{\be}{\begin{equation}}
\newcommand{\en}{\end{equation}}
\newcommand{\bea}{\begin{eqnarray}}
\newcommand{\ena}{\end{eqnarray}}
\newcommand{\beano}{\begin{eqnarray*}}
\newcommand{\enano}{\end{eqnarray*}}

\newcommand{\A}{{\cal A}}

\renewcommand{\l}{\langle}
\renewcommand{\r}{\rangle}
\newcommand{\pin}[2]{\l#1 , #2\r}

\newcommand{\1}{1 \!\! 1}

\newcommand{\Hil}{\mc H}

\catcode `\@=11 \@addtoreset{equation}{section}

\catcode `\@=12
\begin{document}

\begin{center}
{\Large \textbf{Phase transitions, KMS-condition and Decision Making}} \vspace{2cm%
}\\[0pt]

{\large F. Bagarello}
\vspace{3mm}\\[0pt]
Dipartimento di Ingegneria,\\[0pt]
Universit\`{a} di Palermo, I - 90128 Palermo,\\
and I.N.F.N., Sezione di Catania\\
E-mail: fabio.bagarello@unipa.it\\

\vspace{7mm}

\end{center}

\vspace*{2cm}

\begin{abstract}
\noindent We consider a simple model of interacting agents asked to choose between "yes" and "not" to some given question. The agents are described in terms of spin variables, and they interact according to a mean field Heisenberg model. We discuss under which conditions the agents can come out with a common choice. This is made using, in a social context, the notion of KMS-states and phase transitions. 
\end{abstract}

\vspace*{1cm}


\vfill

\newpage


\section{Introduction}

It is nowadays believed by many authors that quantum tools, and ideas borrowed from quantum mechanics, can be relevant in the understanding of several macroscopic systems. Hundreds of papers have been written which explore these connections, and several monographs, like \cite{baa}-\cite{bagbook2}, where many other references can also be found.

Apart from other applications, many authors have considered the role of {\em quantum} in Decision Making and in social sciences, \cite{agra}-\cite{pothbus}. Many aspects of this connection have been discussed in recent years, with different techniques and strategies. 
The partial list of references above  already gives an idea of the interest around this approach. This is mainly due to its many technical  (but not only) consequences, as well as to the very good results which can be found following this idea.

The use of quantum ideas is just one natural step which continues the long story of applications of ideas, originally born in Physics, to other realms of Science, apparently far from hard Sciences. We can find several applications of this kind to Social Sciences, Decision Making, Politics, Economics, Game Theory, just to cite a few.  Econophysics and Sociophysics are nowadays very active lines of research, see \cite{mantegna,galambook}, and many authors have used, in particular, statistical mechanics methods to deal with systems which are far from ordinary Statistical Mechanics, see \cite{galam2,ortega,spagnolo,valenti} just to cite a few.

As we have already noticed, more recently also quantum ideas appeared in these contexts. We should remark that, in the bibliography cited so far on these applications, \cite{baa}-\cite{pothbus}, there is almost no use of Statistical Mechanics, and of phase transitions in particular.

In recent years, A. Khrennikov introduced the notion of {\em social laser}, \cite{andrei1,andrei2,andrei3}, as a sort of general (quantum-like) mechanism to describe the fact that, in social systems, some {\em small} event can be sufficient to produce a {\em large} result, as it happens for real physical lasers. It is like a sort of phase transition occurring under certain conditions, bringing an originally disordered system into an ordered phase. This is exactly the idea we want to explore quantitatively here, by making use of some tools usually adopted in statistical mechanics, and in operator algebras, \cite{brattrob1}-\cite{sewbook2}. More in details, we will show how to use an Heisenberg spin system in its mean-field approximation to mimic the mechanism of a decision making activity when some special condition on the parameters of the system are satisfied. This is similar to what happens for ordinary ferromagnet materials, for which the various spins align only if the temperature is below some critical value. We will show that something similar happens also for our social system, and with a similar mechanism, but with a rather different interpretation. In this perspective the idea of this paper is similar to that considered in \cite{galam}, see also references therein, which adopt statistical methods in the analysis of a decision making problem. The main difference between \cite{galam} and what is discussed here is in the role of the Hamiltonian, and in its related dynamics. This is because, as we will discuss later, the dynamics will be used, via the KMS\footnote{KMS stands for Kubo-Martin-Schwinger.} condition, to find a critical value of a relevant parameter of the model (the {\em temperature}), whose psychological interpretation (the {\em order} of the system) will be discussed later.

The paper is organized as follows: in the next section we introduce the essential ingredients of the mathematical settings needed in our analysis, suggesting some relevant reading for the interested reader for a better comprehension of some details which are not particularly relevant for us, here. Section  \ref{sect3} is devoted to the definition of the model, to its interpretation, and to the analysis of the results in connection with our original problem, which is essentially the following: we have a group of $N$ agents, mutually interacting, which should come out with some binary decision. In general, for instance if the agents are different and if the interaction is weak, there is no particular reason to imagine that they reach a shared collective decision, at least if the agents form a {\em closed system}, i.e. they do not communicate with the external world. On the contrary, it is easy to imagine that such a decision could be reached if the interaction is strong enough or if we open the system to the environment. This is essentially what we will conclude in Section \ref{sect3}, using methods and ideas borrowed by algebraic statistical mechanics and quantum dynamics.  Section \ref{sect4} contains our conclusions.

\section{Introducing the framework}\label{sect2}

Let $\Sc$ be a group of $N$ agents who have to take a decision. The details of this operation will be given in Section \ref{sect3}. Here we just mention that the only possible choices for $\Sc$ are "yes" and "no". Hence, it is natural to imagine that each agent $\tau_j$, $j=1,2,\ldots,N$, is described by a linear combination of two orthogonal vectors, one representing the choice "yes", $e_j^+=\left(
\begin{array}{c}
	1 \\
	0 \\
\end{array}
\right)$, and a second vector, $e_j^-=\left(
\begin{array}{c}
	0 \\
	1 \\
\end{array}
\right)$, corresponding to the choice "no". These two vectors  are an orthonormal (o.n.) basis in the Hilbert space $\Hil_j=\mathbb{C}^2$, endowed with its standard scalar product $\pin{.}{.}_j$. A general vector describing $\tau_j$ can therefore be written as
\be
\psi_j=a_j  e_j^++b_j e_j^-,
\label{21}\en
with $|a_j|^2+|b_j|^2=1$, to ensure the normalization of the vector. Here, as before,  $j=1,2,\ldots,N$. We introduce the following Pauli matrices
$$
\sigma_j^3=\left(
\begin{array}{cc}
	1 & 0 \\
	0 & -1\\
\end{array}
\right), \qquad \sigma_j^+=\left(
\begin{array}{cc}
	0 & 1 \\
	0 & 0\\
\end{array}
\right), \qquad \sigma_j^-=\left(
\begin{array}{cc}
	0 & 0 \\
	1 & 0\\
\end{array}
\right). 
$$
We observe that, $\forall j$, the vectors $e_j^\alpha$ and the matrices $\sigma_j^\beta$, $\alpha=\pm$ and $\beta=3,\pm$, are all copies of the same vectors and matrices. That's why we don't use $j$ also in the explicit expressions of these objects, writing, for instance, $\left(
\begin{array}{c}
	1 \\
	0 \\
\end{array}
\right)_j$ rather than $\left(
\begin{array}{c}
	1 \\
	0 \\
\end{array}
\right)$. This is useful to simplify the notation, when possible.

It is clear that 
\be
\sigma_j^3e_j^{\pm}=\pm e_j^{\pm}, \qquad \sigma_j^\pm e_j^{\mp}=e_j^{\pm}, \qquad \sigma_j^\pm e_j^{\pm}=0.
\label{21b}\en
Because of these formulas, we can consider $\sigma_j^3$ as an operator {\em measuring} the decision of $\tau_j$, according to its eigenvalue: if $\tau_j$ is choosing "yes", then it is described by $e_j^+$, which is an eigenstate of $\sigma_j^3$ with eigenvalue $+1$. Similarly, if $\tau_j$ is choosing "no", then it is described by $e_j^-$, which is also an eigenstate of $\sigma_j^3$, but with eigenvalue $-1$. Of course, $\sigma_j^\pm$ are operators which modify the original attitude of $\tau_j$, according to (\ref{21b}). For instance, after the action of $\sigma_j^+$ on the agent $\tau_j$ originally willing to choose "no", $\tau_j$ is moved to $e_j^+$: he now is going to vote "yes". However, if somebody insists too much, then it is like acting more than once with $\sigma_j^+$ (or with $\sigma_j^-$), and the result is that $\tau_j$ is moved to the zero vector!

Out of $\sigma_j^\pm$ we can also introduce the other two Hermitian Pauli matrices
$$
\sigma_j^1=\sigma_j^++\sigma_j^-=\left(
\begin{array}{cc}
	0 & 1 \\
	1 & 0\\
\end{array}
\right),\qquad \sigma_j^2=i(\sigma_j^--\sigma_j^+)=\left(
\begin{array}{cc}
	0 & -i \\
	i & 0\\
\end{array}
\right), 
$$
and the following rules are easily checked:
\be
(\sigma_j^\pm)^2=0, \qquad [\sigma_j^\pm,\sigma_j^3]=\mp \,2\sigma_j^\pm\qquad [\sigma_j^+,\sigma_j^-]=\sigma_j^3,\qquad \{\sigma_j^+,\sigma_j^-\}=\1,
\label{21c}\en
where $[A,B]=AB-BA$ and $\{A,B\}=AB+BA$ are the commutator and the anti-commutator of the operators $A$ and $B$.

The Hilbert space for $\Sc$ is made of copies of $\mathbb{C}^2$, one for each agent:
$$
\Hil=\otimes_{j=1}^{N}\mathbb{C}_j^2.
$$

An o.n. basis of $\Hil$  consists of tensor products of states $e_j^+$, and $e_j^-$ for various $j$. For instance, the state describing a situation in which the first $L_1$  agents of $\Sc$ votes "yes" and the remaining $L_2$ agents vote "no", $L_1+L_2=N$,  is 
$$
\Phi=(e_1^+\otimes\cdots\otimes e_{L_1}^+)\otimes(e_1^-\otimes\cdots\otimes e_{L_2}^-).
$$
Of course the dimensionality of $\Hil$ increases with $N$. In fact we have $\dim(\Hil)=2^N$. An operator $X_1$ acting, for instance, on $\mathbb{C}_1^2$, is identified with the tensor product $X_1\otimes \1_2\otimes\cdots\1_N$, i.e. the tensor product of $X_1$ with $N-1$ copies of the identity operator $\1$, acting on all the other single-agent Hilbert spaces. In this way we can relate $X_1$ with a (bounded) operator on $\Hil$. The set of all the linear (bounded) operators on $\Hil$ defines an algebra $\A$, which we can call, adopting a standard nomenclature, {\em the algebra of the observables}.
The vector $\Phi$ above, and all other normalized vectors on $\Hil$, defines a state on $\A$, \cite{brattrob1,sewbook1}, as  follows: $\omega_\Phi(A)=\pin{\Phi}{A\,\Phi}$, $A\in\A$. Here $\pin{.}{.}$ is the scalar product on $\Hil$,
$$
\pin{f_1\otimes\cdots\otimes f_N}{g_1\otimes\cdots\otimes g_N}=\prod_{j=1}^{N}\pin{f_j}{g_j}_j,
$$
$f_j,g_j\in\mathbb{C}_j$.
The state $\omega_\Phi$ is only one among all the possible states over $\A$. Other examples of states, i.e. of positive normalized functionals on $\A$, are Gibbs and KMS states, \cite{brattrob2,sewbook1}. These latter are particularly important because of their physical interpretation: they are both equilibrium states at a non zero temperature, while $\omega_\Phi$ typically describe a zero-temperature situation. The difference between Gibbs and KMS states is that, usually, Gibbs states are used for finite dimensional systems, while KMS states are their counterpart for systems with an infinite number of degrees of freedom. Both Gibbs and KMS states satisfy the so-called {KMS-condition}, which we give here in a simplified version. We refer to \cite{brattrob2,sewbook1,sewbook2} for more details, and for a deeper mathematical analysis, of KMS states. The KMS condition is
\be
\omega(AB)=\omega(BA(i\beta)),
\label{22}\en
where $A,B\in\A$, $A(t)$ is the time evolution of $A$ and $A(i\beta)$ is the analytic continuation of $A(t)$ computed in $t=i\beta$, $\beta=\frac{1}{T}$, the inverse temperature (in units in which the Boltzmann constant is equal to one). In the Heisenberg picture, if $H$ is the time-independent Hamiltonian of the system, $A(t)$ is the following operator: $A(t)=e^{iHt}Ae^{-iHt}$. Also, $\omega$ can be written as
\be
\omega(A)=tr(\rho A), \qquad \rho=\frac{e^{-\beta H}}{tr(e^{\beta H})}.
\label{23}\en
Here $tr(X)$ is the trace of the operator $X$, and the KMS condition can be deduced (at least formally) using the cyclicity of the trace, together with the explicit expressions of $\rho$ and $A(t)$. 

\vspace{2mm}

{\bf Remark:--} In many situations a mathematically rigorous approach would imply a preliminary analysis of a reduced system $\Sc_V$, related to the original one, $\Sc$, where $V$  could be the finite volume of the system, or the, again finite, number of particles, or any other possible way to introduce a cutoff in the system. For this finite system, the Hamiltonian $H_V$ can be defined in a rigorous way and then the dynamics can be deduced, $\alpha_V^t(X)=e^{iH_Vt}Xe^{-iH_Vt}$, for all $X\in\A$. Then, to come back to the original system, one has to take the limit $V\rightarrow\infty$. In the literature this is usually called {\em thermodynamical limit}, see \cite{brattrob1}-\cite{bagmaster} and references therein, and requires the use of suitable topologies on $\A$. In other words, while $\Sc_V$ admits, in particular, a well defined {\em quantum energy}, $H_V$, the original, infinitely extended, system does not\footnote{In many cases $H_V$ is a finite sum of contributions, while $H$ would be a non necessarily convergent series. This is, for instance, the difference between $h_N$ in (\ref{33}) and its limit $h_\infty$, which is not convergent, if $J_{i,j}$ and $p_{i,j}$ are not properly chosen.}. In what follows, we will try to simplify the problem as much as possible, just focusing on the results rather than on the mathematical details. For instance, we will not discuss here, since this is not relevant for our model (but could be very relevant for others!) what happens for (\ref{23}) and for the formula $A(t)=e^{iHt}Ae^{-iHt}$ when $H$ is unbounded, as often happens in quantum mechanical systems, or when $H$ is not self-adjoint, which is true for some gain and loss system, \cite{benbook}.

\section{The model}\label{sect3}

We are now ready to introduce the details of the system $\Sc$ we are interested in, and the model we will use to deduce some of the features of the system. As we have briefly anticipated in the previous section,  $\Sc$ is a group of $N$ agents which have to decide between "yes" and "no". The decision should take place after the agents interact among them. In this perspective, the system is {\em closed}: no input from outside will be considered. However, we will also comment on this possibility briefly in the following.

The single agent is described in terms of Pauli matrices: in view of what we have seen in the previous section, we have {\em changing opinion} operators, $\sigma_j^\pm$, and the {\em counting opinion} operator $\sigma_j^3$. The full system $\Sc$ is made by $N$ agents, and the description lives in $\Hil=\otimes_{j=1}^{N}\mathbb{C}_j^2$, as discussed in Section \ref{sect2}. As often in these cases, \cite{bagbook,bagbook2}, the dynamics of $\Sc$ is given in terms of an Hamiltonian operator $H_N$ which is constructed in order to describe the (possibly) more relevant effects occurring in $\Sc$. The first effect we want to consider is the following:
\be
H_{coop}^{(N)}=\sum_{i,j=1}^{N}\,J_{i,j}\sigma_i^3\sigma_j^3,
\label{31}\en
where the label {\em coop} stands for {\em cooperative}. The reason for this name is the following: suppose that the {\em two-body potential} $J_{i,j}$ is positive\footnote{For instance, we could think that $J_{i,j}=\frac{1}{|i-j|}$, which describes a Coulomb-like interaction.}, for all $i,j=1,2,\ldots,N$. Then the contribution to the eigenvalues of $H_{coop}^{(N)}$ is positive if $\tau_i$ and $\tau_j$ are both in the up state, $e_i^+$ and $e_j^+$, or both in the down state, $e_i^-$ and $e_j^-$. On the other hand, if $J_{i,j}<0$, then the contribution of $J_{i,j}\sigma_i^3\sigma_j^3$ to the eigenvalues of $H_{coop}^{(N)}$ is positive only when $\tau_i$ and $\tau_j$ are in opposite states,  $e_i^+$ and $e_j^-$, or  $e_i^-$ and $e_j^+$. In what follows, this latter will be the relevant situation for us. This means that (again, assuming that $J_{i,j}<0$) $H_{coop}^{(N)}$ will be {\em minimized} if the various interacting agents are in the same state since, when this is the case, the corresponding eigenvalue of $H_{coop}^{(N)}$ is negative, and it is the largest possible one in absolute value. Then the spins are "energetically forced" to be all parallel (up or down, it does not matter). This is why we have used here the word "cooperative" in (\ref{31}).

A second natural term to consider for the dynamics of $\Sc$, and therefore in its Hamiltonian, is the following
\be
H_{opp}^{(N)}=\sum_{i,j=1}^{N}\,p_{i,j}\left(\sigma_i^+\sigma_j^-+\sigma_i^-\sigma_j^+\right).
\label{32}\en
The meaning of this term is more psychological than energetic: it describes a situation in which, during an interaction between $\tau_i$ and $\tau_j$, the two agents tend to act in an opposite way: if $\tau_i$ moves from a "no" to a "yes" decision (because of $\sigma_i^+$), $\tau_j$ moves in the opposite direction, because of $\sigma_j^-$. Hence, $H_{opp}^{(N)}$ describes an effect which is the opposite with respect to that produced by (\ref{31}). Of course, $p_{i,j}$ gives the {\em strength} of the mutual tendency of the agents to behave differently. Is a sort of {\em mutual antipathy} or {\em mutual disagreement}.

Putting the two contributions together, we introduce the Hamiltonian
\be
h_N=H_{coop}^{(N)}+H_{opp}^{(N)}=\sum_{i,j=1}^{N}\,J_{i,j}\sigma_i^3\sigma_j^3+\sum_{i,j=1}^{N}\,p_{i,j}\left(\sigma_i^+\sigma_j^-+\sigma_i^-\sigma_j^+\right),
\label{33}\en
which will be the starting point of our analysis.
\vspace{2mm}

{\bf Remark:--} we could enrich this model adding to $h_N$ a third (and many other!) contribution leading to a preferred direction for the various spin operators, by suggesting to the various agents to choose up or down. This extra term is easily constructed:
\be
H_{ext}^{(N)}=B\sum_{i=1}^{N}\,\sigma_i^3,
\label{34}\en
where $B\in\mathbb{R}$. Once more, we use an energetic understanding of its meaning: if $B>0$, then the lowest eigenvalue of $H_{ext}^{(N)}$ is obtained by a vector with all down vectors, $e_1^-\otimes\cdots\otimes e_{N}^-$, and its corresponding eigenvalue is clearly $-NB$. Analogously, if we take $B<0$, then the lowest eigenvalue of $H_{ext}^{(N)}$ is $NB$, with corresponding eigenvector  $e_1^+\otimes\cdots\otimes e_{N}^+$. A possible interpretation of  $H_{ext}^{(N)}$ is that it can be used to model the presence of some information coming, for instance, from what is surrounding $\Sc$, and which is uniformly diffused in $\Sc$, and contribute to the final decision of the agents. However, in this paper we will not include this term in $h_N$ since, as already stated, we only want to consider what is going in our closed system. Moreover, and possibly more important, the presence of $B$ forces the spins to be aligned, and this makes our analysis on the competitive effects of $H_{coop}^{(N)}$ and $H_{opp}^{(N)}$ less interesting, removing some freedom to the agents.

\vspace{2mm}

Since $\sigma_i^+\sigma_j^-+\sigma_i^-\sigma_j^+=\frac{1}{2}(\sigma_i^1\sigma_j^1+\sigma_i^2\sigma_j^2)$, the Hamiltonian $h_N$ can be rewritten only in terms of $\sigma_j^\alpha$, with $\alpha=1,2,3$. Moreover, let us assume that both $J_{i,j}$ and $p_{i,j}$ only depend on the number of the agents $N$, and not on the nature of the single $\tau_j$. Hence, as it is often discussed in statistical mechanics, \cite{ruelle,baxter}, we can consider the following {\em mean field approximation}:
$$
p_{i,j}\rightarrow \frac{p}{N}, \qquad J_{i,j}\rightarrow \frac{J}{N}.
$$
If we further assume that $J$ and $p$ are related, and that in particular that $p=2J$, we can replace  $h_N$ with its mean field approximation $H_N$:
\be
H_N=\frac{J}{N}\sum_{i,j=1}^{N}\sum_{\alpha=1}^3\sigma_i^\alpha\sigma_j^\alpha,
\label{35}\en
which is known as the mean-field Heisenberg model, \cite{ruelle}. The existence of the dynamics for $H_N$, and of the thermodynamical limit of this model ($N\rightarrow\infty$), has been discussed at length in, for instance, \cite{bagmorc,bagmaster}, where it is essentially proved what is listed below (focusing only on those aspects which are relevant for us, here):

\begin{enumerate}

\item the operator $\sigma_N^\alpha=\frac{1}{N}\sum_{i=1}^{N}\sigma_i^\alpha$ converges (in the strong topology restricted to a suitable set of state) to an operator $\sigma_\infty^\alpha$ which commutes with all the $\sigma_k^\beta$, $\forall k$ and $\forall\beta$, and therefore belongs to the center of $\A$.

\item $H_N$ generates a time evolution of each $\sigma_\alpha^i$, $\alpha_N^t(\sigma_\alpha^i)=e^{iH_Nt}\sigma_\alpha^ie^{-iH_Nt}$, which converges (again, in the strong topology) 
\be
\alpha^t(\sigma^\alpha_i)=\cos^2(Ft)\sigma^\alpha_i+\frac{i}{F}\sin(Ft)\cos(Ft)[\underline{F}\cdot\underline\sigma_i,\sigma^\alpha_i]+\frac{1}{F^2}\sin^2(Ft)(\underline{F}\cdot\underline\sigma_i) \,\sigma^\alpha_i (\underline{F}\cdot\underline\sigma_i),
\label{36}\en
	where\footnote{Here we adopt the folowing standard notation: $\underline v=(v^1,v^2,v^3)$.} $\underline{F}=J\underline \sigma_\infty$ and $F=\|\underline{F}\|=|J|\|\underline \sigma_\infty\|$. We observe that (\ref{36}) is well defined for all $F\neq0$, and that it can also be extended continuously to the case $F=0$. 
	
	\item If we are now interested in the expression of $\alpha^t(\sigma^\alpha_i)$ in a representation $\pi_\omega$, arising as the GNS-representation\footnote{The mathematical details of this construction are not very relevant for us, and can be found in \cite{bagmorc}. The relevant aspect here is that different representations correspond to different physics for the same system, \cite{brattrob1,brattrob2}. For instance, the same formula (\ref{36}) describes ferromagnetic and anti-ferromagnetic systems. But these systems are recovered using different GNS-representations. For completeness, we recall that GNS stands for Gelfand, Naimark and Segal.} of a state $\omega$ over $\A$, formula (\ref{36}) produces
	\be
	\alpha_\pi^t(\sigma^\alpha_i)=\cos^2(ft)\sigma^\alpha_i+\frac{i}{f}\sin(ft)\cos(ft)[\underline{f}\cdot\underline\sigma_i,\sigma^\alpha_i]+\frac{1}{f^2}\sin^2(ft)(\underline{f}\cdot\underline\sigma_i) \,\sigma^\alpha_i (\underline{f}\cdot\underline\sigma_i),
	\label{37}\en
	where $\underline f=\pi_\omega(\underline F)=J\underline m$, $\underline m=\pi_\omega(\underline \sigma_\infty)$. Here and in (\ref{36}) $\alpha=1,2,3$, while $i=1,2,3,\ldots,N$.
	
	\item The effective dynamics $\alpha_\pi^t$ can be deduced by the following effective, representation-dependent, Hamiltonian
	$$
	H_\pi=\underline f\cdot\sum_{i=1}^\infty \underline \sigma_i.
	$$
	Of course, using $H_\pi$ rather than $H_N$ simplifies the analysis of the dynamics of $\Sc$ quite a bit, \cite{bagmorc,bagmaster}, and, for this reason, and since it does not change the form of the dynamics, $H_\pi$ will be used in the rest of this paper.
	
\end{enumerate}

Because of our interpretation, $m$ has an interesting meaning for us, as in the standard Heisenberg model, where it is essentially the magnetization of the spin system. Here $m$ is a measure of how close our agents are to a common decision: if $m\simeq 0$, their decisions are so different that their mean value is close to zero. In fact we can write $\underline m=\lim_{N,\infty}\frac{1}{N}\sum_{i=1}^N\omega(\underline \sigma_i)$, so that $m=\|\underline m\|$ is essentially a measure of the mean value of the choices of each agent. On the other hand, the closer $m$ is to one, the more all the agents (or most of them) are reaching the same conclusion, even if we don't know which one is the preferred, between "yes" and "no". In the first case ($m\simeq0$) the effect of $H_{opp}^{(N)}$ wins over $H_{coop}^{(N)}$, while when $m\simeq1$, the opposite happens.

The next step of our analysis consists in using the KMS identity (\ref{22}) with formula (\ref{37}). Due to the properties of the GNS representation, it is known that  $\underline m=\pi_\omega(\underline \sigma_\infty)=\omega(\underline \sigma_i)$, $\forall i$. We see that the dependence of the agent index $i$ disappears. This is because all our agents are {\em equivalent} in the treatment discussed here, as the analytic expression of $H_N$ in (\ref{35}) clearly shows. Hence, after some algebra, using (\ref{21c}) and putting $A=B=\sigma_1^i$, we can rewrite the equality in (\ref{22}), $\omega(\sigma_1^i\sigma_1^i)=\omega(\sigma_1^i\sigma_1^i(i\beta))$, as follows:
\be
\sinh(Jm\beta)(m_2^2+m_3^2)\left[\frac{1}{m}\cosh(Jm\beta)+\frac{1}{m^2}\sinh(Jm\beta)\right]=0.
\label{38}\en
Notice that, despite of the appearance of $m=\|\underline m\|$ in the denominators, formula (\ref{38}) is well defined also in the limit $m\rightarrow0$. As discussed before, a solution $m=0$ can be seen as the incapacity of the agents to find an agreement on the original "yes"-"no" question. Much more interesting, for us, is to find if and under which condition a non zero solution of (\ref{38}) does exist. Of course, at a first sight, one can imagine that any vector of the form $\underline m_a=(m_1,0,0)$ is such a solution, if $m_1\neq0$. And, in fact, it is clear that 
$\underline m_a$ solves (\ref{38}), and that it is non-zero. However, $\underline m_a$ is not a solution of the KMS condition when we replace $\sigma_1^i$ with, say, $\sigma_2^i$. In this case, a non zero solution of the related KMS condition would be, as it is easily understood, $\underline m_b=(0,m_2,0)$, $m_2\neq0$. In the same way, $\underline m_c=(0,0,m_3)$, $m_3\neq0$, would be a non zero solution of the KMS condition with $A=B=\sigma_3^i$. Since the KMS condition should be satisfied for any choice of $A$ and $B$ in $\A$, it is clear that the only possible common solution for these three different choices of $A$ and $B$ is $\underline m_0=\underline 0$. This preliminary analysis implies that, if we are only interested in non zero solutions, the term $\sinh(Jm\beta)(m_2^2+m_3^2)$ in (\ref{38}) is not so important. What is important, on the contrary, is the other part, which can be safely rewritten as
\be
\tanh(Jm\beta)=-m,
\label{39}\en
which is a very well known condition one often meets when dealing with many body systems, see e.g. \cite{ruelle,baxter,bagmaster} and references therein. Here $J$ is the parameter we introduced when going from the Hamiltonian $h_N$ to its mean field version $H_N$ in (\ref{35}). We also recall that, in our derivation, $J$ was also assumed to be proportional to $p$, and that $J$ needs to be taken negative if we want the energy of $H_{coop}^{(N)}$ to be minimized by parallel spins (or, in our present situation, by agents sharing the same idea). This is in agreement with the analysis which follows, which we include here mainly for those readers which are not very familiar with statistical mechanics. Let us introduce the function $\Phi(m)=m+\tanh(Jm\beta)$, in which $J$ and $\beta$ are seen as parameters. It is clear that $m\geq0$ necessarily, since $m$ is the norm of a vector. It is also clear that, since $\tanh(x)\in]-1,1[$ for all $x\in\mathbb{R}$, $\Phi(0)=0$ and $\lim_{m,\infty}\Phi(m)=\infty$. Then, if $\Phi'(0)<0$, we are sure that the function $\Phi(m)$ acquires negative values in the right neighbourhood of $m=0$. But $\Phi(m)$ is a continuous function. Hence a value $m_c\neq0$ must exists such that $\Phi(m_c)=0$. This $m_c$ is, of course, the non zero solution of (\ref{39}) we were looking for. It is clear that this simple argument is not sufficient to conclude if more than a single non trivial zero of $\Phi(m)$ could exist. But, for what is relevant for us, one solution is enough.

We have $\Phi'(m)=1+\frac{J\beta}{\cosh^2(J\beta m)}$, and since $\cosh(0)=1$ it follows that $\Phi'(0)=1+J\beta$, which is negative only if $J\beta<-1$. Since $\beta=\frac{1}{T}$ is always positive, the inequality can only be satisfied if $J$ is negative, which appears to be in agreement with our working condition. Now, calling $T_c=-J$, we can conclude that a non zero solution of (\ref{39}) does exist if $T<T_c$, i.e., for temperatures less than a critical value which is proportional (with a minus sign) to the strength with which the agents are interacting. Before moving to the interpretation of this result in our specific context, it is also interesting to observe that, when $-J\beta$ increases,  $m_c$ approaches one. This can be seen numerically, but also directly from (\ref{39}): if $m\neq0$, then when $-J\beta\rightarrow+\infty$ also $-Jm\beta\rightarrow+\infty$, and therefore $\tanh(Jm\beta)\rightarrow-1$, which implies that $m\rightarrow1$, necessarily, from the equality in (\ref{39}).

These results suggest the following interpretation: the various agents can cooperatively come out with a common  decision only in presence of some {\em lack of disorder} in $\Sc$. In other words, we can see at $T$ (the temperature, in its original interpretation) as a measure of the order in the system, as in a standard thermodynamical interpretation: the higher the temperature, the higher the disorder. On the opposite side, when the temperature decreases the order increases, and then it is possible to get a common shared decision. Hence we can look at $T=\beta^{-1}$ in $\omega$ as a sort of {\em order parameter}, and at $J=-T_c$ as the critical value of this parameter measuring a phase transition from order to disorder, or vice-versa. A small change may produce large effects, like in the social laser approach, \cite{andrei1}-\cite{andrei3}. Moreover, since we can rewrite $-J\beta=\frac{T_c}{T}$, our previous analysis also shows that, if $T\ll T_c$, $m$ approaches one. This is in agreement with our interpretation: the lower the {\em temperature} of $\Sc$, the closer the result to $\pm 1$: the degree of disorder decreases and all the agents (or a large part of them) share the same decision.

\section{Conclusions}\label{sect4}

This is a first application, or one of the few existing so far in our knowledge, of methods of algebraic quantum dynamics and algebraic statistical mechanics to social systems, and more specifically to decision making. A possible role of KMS states in similar contexts, and in particular to finance, was first proposed in \cite{bag1}, but never explored after that first attempt. Here, having in mind a rather abstract problem in decision making, we discuss how the KMS condition can be used to describe a sort of phase transition, from "no shared collective decision" to "we share a decision!". This phase transition has been found for a mean field spin model, under a suitable condition on the parameters of the Hamiltonian assumed for the description of the dynamics of the system and of the state describing $\Sc$.

In other words, we suggest a possible { quantitative} approach which produces a sort of collective decision, using a set of minimal (although mathematically not so simple) ingredients and techniques which, we believe,  can be relevant for further studies in the area of decision making. This is quite different from what can be found in some literature on the same topics, in which there are some ideas on why a common decision can be taken, but there is no {\em dynamical reason} for that. Our approach is different: we focus on the dynamics of the system and, therefore, we suggest an explicit form of the interaction between the agents producing an {\em energy-like operator} (the Hamiltonian) which, in turn, is the source of the dynamics of the system. Then we use this dynamics to show that a phase transition can be deduced, and this phase transition occurs only when we are below a certain critical parameter, measuring the order/disorder of the system. Since the system is rather general, our approach suggests a general mechanism which can be adapted to different explicit systems to check if and when a phase transition may occur. The fact that here a phase transition corresponds to a collective decision is just because of the particular system we have considered in this paper, but the same approach (changing, of course, the Hamiltonian in order to implement these differences) could work also for quite different systems.

As already mentioned, the model considered here can be seen as an explicit appearance of a social laser, driven by just a pair of parameters appearing in the definition of the model itself. We should also stress once more that, under some aspects, the framework proposed in this paper is connected to that in \cite{galambook,ortega,spagnolo,valenti}, since all these approaches adopt techniques coming from Physics, but with some substantial differences.

It could be worth mentioning that an evident difference exists between the model proposed here and older ones, proposed by us and our group, more focused on the dynamical behaviour of a given system. In particular, here we have used spin variables, while in many previous applications we have rather used fermionic or bosonic operators. Our different choice here is useful to easily translate some results originally deduced for a spin chain which is very well known in statistical mechanics, the Heisenberg model, to our totally different context, the set of our agents.  Of course, we could rewrite the model, and the Hamiltonian $H_N$ in particular, in terms of fermionic matrices, recovering the general framework adopted in several previous applications, \cite{bagbook,bagbook2}, but this is not so relevant for us, here.

The next step is, of course, a deeper understanding of this kind of phase transitions in macroscopic systems, and in particular of the {\em real} role of the KMS condition. This understanding will be easier in presence of other models which, hopefully, should clarify the relevance of our approach in similar problems.

\section*{Acknowledgements}

The author acknowledges partial financial support from Palermo University (via FFR2021 "Bagarello") and from G.N.F.M. of the INdAM.

\section*{Data accessibility statement}

This work does not have any experimental data.

%
%
%
%
%

\section*{Funding statement}

The author acknowledges partial financial support from Palermo University, via FFR2021 "Bagarello".

\end{document}